# Graphene-Enhanced Hybrid Phase Change Materials for Thermal Management of Li-Ion Batteries

Pradyumna Goli, Stanislav Legedza, Aditya Dhar, Ruben Salgado, Jacqueline Renteria and Alexander A. Balandin[*]

Nano-Device Laboratory, Department of Electrical Engineering and Materials Science and Engineering Program, Bourns College of Engineering, University of California – Riverside, Riverside, CA 92521 USA

Li-ion batteries are crucial components for progress in mobile communications and transport technologies. However, Li-ion batteries suffer from strong self-heating, which limits their life-time and creates reliability and environmental problems. Here we show that thermal management and the reliability of Li-ion batteries can be drastically improved using hybrid phase change material with graphene fillers. Conventional thermal management of batteries relies on the latent heat stored in the phase change material as its phase changes over a small temperature range, thereby reducing the temperature rise inside the battery. Incorporation of graphene to the hydrocarbon-based phase change material allows one to increase its thermal conductivity by more than *two orders of magnitude* while preserving its latent heat storage ability. A combination of the sensible and latent heat storage together with the improved heat conduction outside of the battery pack leads to a significant decrease in the temperature rise inside a typical Li-ion battery pack. The described combined heat storage – heat conduction approach can lead to a transformative change in thermal management of Li-ion and other types of batteries.

---

[*] Corresponding author (A.A.B.): balandin@ee.ucr.edu

Development of Lithium-ion (Li-ion) batteries enabled progress in mobile communications, consumer electronics, automotive and aerospace industries. Li-ion batteries are an essential part of the hybrid electric vehicles (HEV) owing to their high energy densities and low weight-to-volume ratios [1]. One of the most significant factors negatively affecting Li-ion battery performance is a temperature rise beyond the normal operating range. If overheated due to short-circuiting or fast charging/discharging processes the Li-ion battery can suffer thermal runaway, cell rupture or even explosion [2]. A fire in the Li-ion battery results in the emission of dense irritating smoke which could present a serious health and environmental risk [2-3]. Combining multiple Li-ion cells close together in a battery pack in order to provide higher electric power makes the thermal management of the batteries even more challenging. The severity of the potential thermal issues with the battery packs is exemplified by a recent incident with the overheating and fire in the batteries on-board the Boeing 787 Dreamliner [4].

A common approach for thermal management of Li-ion battery packs is based on the utilization of phase-change materials (PCM). The latent heat stored in PCM, as its phase changes over a small temperature range, allows one to reduce the temperature rise inside the battery [5-7]. By varying the chemical composition of PCM one can tune its melting point and the temperature range in which it can operate as a heat absorber. It is important to note that common PCMs are characterized by very low thermal conductivity, $K$, with typical values in the range of $0.17 - 0.35$ W/mK at room temperature (RT) [8]. For comparison, the RT thermal conductivity of silicon and copper are ~145 W/mK and ~381 W/mK, respectively. PCMs store heat from the batteries rather than transfer it away from the battery pack. The use of PCM in battery cells also serves the purpose of buffering the Li-ion cell from extreme fluctuations in ambient temperature. This is a different approach from what is used in the thermal management of computer chips. In order to reduce the temperature rise in a computer chip one uses thin layers of thermal interface materials (TIMs) or heat spreaders that transfer heat from the chips to heat sinks and outside packaging [9-11]. The thermal conductivity of TIMs is in the range of $1 - 25$ W/mK while that of solid graphite-based heat spreaders can be on the order of $10^3$ W/mK [12].

Here we show that these two different approaches for thermal management can be combined via introduction of the hybrid PCM with graphene acting as filler for increased thermal conductivity.

Graphene is known to have extremely high intrinsic thermal conductivity [13-14] and form excellent binding with a variety of matrix materials [11, 15-16]. The graphene-enhanced hybrid PCM reveals thermal conductivity that is *two orders of magnitude higher* than that of conventional PCM while preserving its latent heat storage ability. Utilization of the hybrid PCM results in substantial decrease of the temperature rise inside battery packs as demonstrated under realistic conditions.

In order to demonstrate possible enhancement of thermal properties with graphene we selected paraffin wax (IGI-1260) as the base PCM. Paraffinic hydrocarbons, or paraffins, are straight-chain or branching saturated organic compounds with the composition $C_nH_{2n+2}$. The term paraffin wax refers to mixtures of various hydrocarbon groups, particularly paraffins and cycloalkanes that are solid at ambient temperature [17]. Paraffin waxes are commonly used PCMs owing to their availability, chemical stability, and durability to cycling. Paraffin has a high latent heat of fusion (200 – 250 kJ/kg) and a range of melting points suitable for thermal control of batteries and portable electronics. The IGI-1260 paraffin wax has relatively high melting and boiling points of $T_M \sim 70$ $^o$C and $T_B \sim 90$ $^o$C, respectively. It consists of $C_{34}$-$C_{35}$ hydrocarbons, which are mainly composed of n-alkanes [17]. The long hydrocarbon chains are responsible for its high density and melting point. When heated the IGI-1260 wax absorbs the heat to break the longer hydrocarbon chains into smaller ones.

The hybrid graphene-PCM composites were prepared by dispersing a solution of the liquid-phase exfoliated (LPE) graphene and few-layer graphene (FLG) in the paraffin wax at 70 $^o$C followed by the high-shear mixing on a hot plate (Corning PC-620D) with a magnetic stirrer. The preparation temperature was selected to avoid oxidation of the paraffin wax with formation of peroxide and water. The hybrid graphene-PCM was put in molds and allowed to solidify at RT under controlled humidity conditions. Three types of LPE graphene were used for the filler. For low loading fractions up to 1%, we used graphene solution with the average thickness of one monolayer (0.35 nm) and the lateral size distribution in the range from 150 to 3000 nm with 550 nm average size. We refer to this material, which is predominantly single-layer graphene, as graphene filler type A. For high loading fractions up to 20%, we used two other types of FLG. The graphene filler type B had an average FLG flake thickness of ~ 1 nm, which constitutes about 3 atomic planes with an average lateral dimension of ~10 μm. The graphene filler type C had an average flake thickness of 8 nm, which constitutes 20-30 atomic planes with a lateral size in the range of 150 and 3000 nm with ~550-nm average. Figure 1 (a) shows an optical image of the resulting molded disks of the hybrid graphene-PCM composite with the graphene-FLG loading fraction varying from 0.5 to 20 wt. %. The color of the disks changes from white to black as the fraction of graphene increases. It was observed that the addition of graphene leads to some reduction of $T_M$. The uniformity of graphene-FLG dispersion in the paraffin matrix was checked with the scanning electron microscopy (SEM) (see Figure 1 (b)).

The incorporation of graphene into paraffin matrix was monitored using micro-Raman spectroscopy (Renishaw In-Via). The vibrational spectra of paraffin are known to have a large number of informative bands that show variations with the change in paraffin's state and composition [18-22]. The measurements were performed in the backscattering configuration under $\lambda = 488$ nm laser excitation. Figure 1 (c) shows Raman spectrum of the hybrid graphene – paraffin wax. The clearly identified vibrational bands are $CH_2$ rocking at ~650 – 850 $cm^{-1}$, C-C skeletal stretching at 1060 $cm^{-1}$ (symmetric) and 1130 $cm^{-1}$ (asymmetric), $CH_2$ twisting at 1300 $cm^{-1}$, $CH_2$ bending at 1440 $cm^{-1}$, and overtones of bending vibrations at above ~2000 $cm^{-1}$. A small peak at 1580 $cm^{-1}$ was identified as graphene's signature G peak. Its intensity is much lower than that of long hydrocarbon chains of paraffin. Graphene's incorporation into paraffin resulted in changes of some paraffin peaks and the appearance of new features in the spectra suggesting modification of the vibrational modes due to the attachment of graphene flakes to the long hydrocarbon chains. Although it is difficult to quantitatively describe the changes, the detailed calibration of Raman spectra with the amount of graphene loading and sample preparation conditions allowed us to achieve a consistent composition of the hybrid graphene-PCM. The changes with the hydrocarbon chains and increasing concentration of $sp^2$-bonded carbon have also been confirmed with X-ray photoelectron spectroscopy (XPS) analysis (Kratos AXIS ULTRADLD). The carbon atoms in the hydrocarbon chains were identified by 284.9 eV peak of a 1s-orbital electron. Figure 1 (d) shows the counts per second of hydrocarbon as a function of graphene concentration in the composites.

The thermal conductivity of the hybrid composites was measured using the transient planar source (TPS) technique (Hot Disk TPS2500) [23]. This method is best suited for the examined class of materials and was previously used for the investigation of thermal properties of other PCM [24] and thermal greases [25]. We calibrated our TPS system by measuring reference samples with known thermal conductivity. We also compared the results of our measurements with those obtained by other experimental techniques such as "laser flash" and "3-omega" [26-27]. The details of the measurement procedures are given in the *Methods* section. Figure 2 (a) presents the measured thermal conductivity of the pristine paraffin wax IGI-1260 and hybrid PCM composites with different graphene- FLG loading. More than ten samples were investigated for each loading fraction to ensure reproducibility. The measured thermal conductivity for the pristine paraffin was $K$=0.25 W/mK, which is in agreement with the literature values. One can see a drastic increase of $K$ in the composites with the addition of graphene-FLG filler. The thermal conductivity of the hybrid graphene-PCM reaches ~15 W/mK at RT with the small 1 wt. % loading fraction. This is a significant increase by a factor of 60. The highest value achieved at 20 wt. % loading was ~45 W/mK, which is more than a *two order magnitude* of enhancement.

The thermal conductivity enhancement factor, $\eta$=$(K-K_m)/K_m$, of about 60 at the 1 wt. % loading fraction is exceptionally high compared with the values reported for either PCMs with fillers [28-30] or TIMs [11, 15-16] ($K$ is the measured thermal conductivity of the composite and $K_m$ is the thermal conductivity of the paraffin matrix). It is unlikely that uniformly dispersed graphene flakes with a lateral size in the range from 150 to 3000 nm form a thermally percolating network at 1 wt. % by themselves. A more probable explanation of the strongly increased thermal conductivity of the composite is easy attachment of hydrocarbon molecules to graphene flakes at the experimentally determined processing temperature. The $C_nH_{2n+2}$ − graphene attachment reduces the thermal interface resistance between the matrix material and filler. Modification of some of the Raman signatures of paraffin after addition of graphene is consistent with this assertion. It was reported [13, 16] that graphene has a much lower thermal Kapitza resistance, $R_B$=$\Delta T/(Q/A)$, with many matrix materials as compared to carbon nanotubes (here $\Delta T$ is temperature differences between two materials forming an interface, $Q$ is the heat flux and $A$ is the surface area).

In the case of paraffin and graphene the thermal coupling between the matrix and filler is likely even stronger than in other matrix-filler combinations. The *ab initio* density function theory calculations and molecular dynamics simulations suggested the possibility of extraordinary enhancement of thermal conductivity in ordered graphene composites with organic matrix where the heat transport is along the direction of the graphene planes: $K/K_m \approx 360$ at graphene loading of 5% [31]. The thermal conductivity in the direction perpendicular to the graphene planes almost does not change, according to the same study [31]. The strong anisotropic increase in the heat conduction was attributed to graphene's planar geometry and strong coupling to the octane molecules resulting in the corresponding decrease in the Kapitza resistance [31-33]. This means that heat carrying phonon modes excited in graphene can couple well to those in organic molecules. Although a direct quantitative comparison between our graphene-paraffin composite and the composite studied in Ref. [31] is not possible one can conclude that even randomly oriented graphene flakes should produce significant increase in the thermal conductivity of composites in agreement with our experiments.

Thermal conductivity of all composites revealed only weak temperature dependence, which is beneficial for PCM practical applications. This weak dependence is expected for disordered materials. Improvement in the thermal management applications of the hybrid graphene-PCM can only be achieved if the increase in the thermal conductivity is achieved without degradation of the latent and sensible heat storage capacity. Possible changes in $T_M$ due to graphene loading should also be adjusted. We performed the specific heat, $C_p$, measurements (NETZSCH) with a set of the samples with the thicknesses of 1 mm − 1.6 mm to ensure that their thermal resistances were much larger than the contact thermal resistances. As a control experiment we measured specific heat of pristine paraffin wax. Figure 2 (b) presents the specific heat data in the examined temperature range. The specific heat for the reference paraffin wax is ~2 kJ/kgK at RT, which is consistent with literature values. Near RT, the specific heat does not change much with the addition of graphene filler. The difference appears in the higher temperature range. The hybrid graphene-PCM has larger specific heat than the reference paraffin. The growth of $C_p$ at the temperature increases above 320 − 330 K is expected. In paraffins, the specific heat starts to increase as temperature approaches $T_M$ and then falls off again [17].

In order to directly prove that the developed hybrid graphene-PCM composites can significantly improve the thermal management of Li-ion batteries we performed the battery testing under realistic conditions. We used six 4-V Li-ion cells with the capacity of 3000 mAh each placed in a standard aluminum battery pack. The measurements were performed with the charger-discharger setup (HYPERION) and the temperature probes (Extech SDL200) that logged temperature for the assigned time intervals. The first two temperature probes were placed inside the battery pack, the third probe was connected to the battery pack shell acting as the heat sink and the fourth probe was used to collect the ambient temperature data. During the measurements the batteries were charging-discharging at 16 A and 5 A, respectively. Details of the experimental setup and testing are provided in the *Supplemental Information*. The first control experiment was performed with pristine paraffin wax, which was melted and poured into the aluminum cylinder containing Li-ion battery cells. Special care was taken to ensure that the wax completely filled the space between the cylinders as in conventional battery designs. The battery pack with paraffin was allowed to cool to RT and then tested through ten charge/discharge cycles. The experiments with the hybrid graphene-PCM followed the same protocol.

Figure 3 (a) shows temperature as a function of time during the charging-discharging cycles for the Li-ion battery pack with IGI-1260 as PCM. One can see that the temperatures of the anode and cathode are higher than that of the outside shell (indicated in the figure as the battery pack temperature). The variation of the ambient T during the measurement explains some background variation in the temperature cycles. The results of the tests of the hybrid graphene-PCM are summarized in Figure 3 (b). One can see that when no PCM was used in the battery pack (the heat dissipates through the air and metal bottom of the pack) the temperature rise inside the battery (sensor attached to anode/cathode) is the highest: $\Delta T \sim 37$ $^o$C. The use of conventional PCM results in the decrease of the temperature rise to ~24 $^o$C. The Li-ion battery pack with the developed hybrid graphene-PCM reveals the lowest temperature rise of ~10 $^o$C during the first cycle. The temperature rise increases to ~16 $^o$C after the third cycle and saturates at this value. The temperature rise for the case of the hybrid graphene-PCM with the larger loading fraction saturates at ~ 13 $^o$C. One should note here that the outside shell (battery pack) made of thin aluminum was not an optimized heat sink. Attachment of the outside shell to a good heat sink would make the improvement in thermal management with the hybrid graphene PCM even more pronounced.

The dependence of $\Delta T$ on the number of cycles, observed in Figure 3 (b), reflects the physical mechanisms behind the cooling action of conventional PCM and the hybrid graphene-PCM. The conventional PCM mostly absorbs the heat from the battery cylinders conducting only its small portion to the battery pack shell. The hybrid graphene-PCM stores and conducts heat simultaneously. This results in lower $\Delta T$ inside the battery pack but also increases the temperature of the outside shell. The increasing temperature of the shell results in some increase in $\Delta T$ inside the battery as well. In order to elucidate this difference in cooling action, in our experiments we intentionally did not connect the outside shell, which constitute the battery pack, to any specially designed heat sink. In practical automotive and aerospace applications one can readily envision a proper thermal connection of the battery packs to the heat sinks, e.g. to the heavy vehicle frame in HEVs. The latter will eliminate or reduce $\Delta T$ of the outside shell further improving thermal management with the hybrid graphene-PCM.

Computer simulation of the passive PCM thermal management systems for Li-ion battery packs is known to give valuable information for materials and system optimization [36-37]. We further analyzed our experimental results via numerical solution of the heat diffusion equation for the specific battery design and measured specific heat and thermal conductivity. Figure 4 shows the schematic of the Li-ion battery pack and the simulated temperature profiles (COMSOL) for the four cases, which corresponded to the conducted experiments. In the case of no PCM between the battery cylinders and the outside shell, the temperature in the cylinders is at its maximum of above 330 K. The outer shell also heats up to ~315 K via conduction through the air. The use of the standard paraffin wax reduces the temperature of the cylinders to around 320 K without heating the outside shell. Thermal management with hybrid graphene PCM results in the lowest temperature of the battery cylinders of ~310 − 315 K with some increase in the temperature of the outside shell. The temperature profile is much more uniform when the hybrid graphene PCM is used. One should note here again that connecting the outside shell to a proper heat sink would improve the performance of the hybrid PCM further.

In conclusion, we demonstrated that the use of graphene and few-layer graphene as fillers in organic phase change material allows one to increase its thermal conductivity by more than two orders of magnitude while preserving its latent heat storage ability. The strong enhancement is achieved via easy binding of graphene flakes to paraffinic hydrocarbons resulting in good thermal coupling. The exceptionally large thermal conductivity of graphene improves the heat conduction ability of paraffins. It was also shown through measurements and computer simulations that improved thermal properties of graphene PCM result in significant temperature rise inside realistic Li-ion battery packs. The described combined heat storage – heat conduction approach may lead to a transformative change in thermal management of batteries.

**METHODS**

**Thermal Conductivity Measurements:** The measurements were performed by sandwiching an electrically insulated flat nickel sensor with the radius 14.61 mm between two identical samples of the same composition. The size of the samples was optimized for the sensor radius. The sensor acted as the heat source and temperature monitor simultaneously. Thermal properties of the material were determined by recording temperature rise as a function of time using the equation $\overline{\Delta T(\tau)} == P \left( \pi^{\frac{3}{2}} r K \right)^{-1} D(\tau)$, where $\tau = (t_m \alpha / r^2)^{1/2}$, $\alpha$ is the thermal diffusivity, $t_m$ is the transient measurement time, $r$ is the radius of the sensor, $P$ is the input heating power, and $D(\tau)$ is the modified Bessel function. The time and the input power were chosen so that the heat flow is within the sample boundaries and the $T$ rise of the sensor is not influenced by the outer boundaries of the sample [23, 34-35].

**Acknowledgements**

The work was supported, in part, by the Semiconductor Research Corporation (SRC) and the Defense Advanced Research Project Agency (DARPA) through FCRP Center for Function Accelerated nanoMaterial Engineering (FAME).

**FUGURE CAPTIONS**

**Figure 1: Hybrid graphene – paraffin phase change material**. (a) Optical image of the PCM samples showing the change in color with increasing graphene content. (b) Scanning electron microscopy image of the hybrid graphene-PCM indicating uniform distribution of the graphene flakes. (c) Raman spectrum of the graphene-paraffin composite. The main bands are indicated in the legends. The graphene G peak is weak compared to hydrocarbon signatures due to its small concentration and smaller scattering cross-section. (d) X-ray photoelectron spectroscopy data for pristine paraffin and the hybrid composites. The intensity of the signal from the hydrocarbon chains, determined via 284.9 eV peak of a 1s-orbital electron, is decreasing with increasing concentration of $sp^2$ carbon of graphene and few-layer graphene.

**Figure 2: Thermal properties of hybrid graphene-PCM.** (a) Thermal conductivity of the graphene – paraffin composites with different graphene loading as the function of temperature. The results for pristine paraffin (IGI-1260) are also shown for comparison. (b) Specific heat of the composites and reference pristine paraffin as the function of temperature.

**Figure 3: Utilization of the hybrid graphene-PCM for thermal management of Li-ion battery pack.** (a) Measured temperature fluctuations inside and outside the battery pack with reference paraffin used as the phase change material. The temperatures are recorded at the battery cylinder cathode (blue), battery cylinder anode (red) and battery pack shell (black). The ambient temperature charge during the measurement is also shown (green). (b) Diagram of the temperature rise inside the Li-ion battery pack during the first ten charging – discharging cycles for the battery pack without PCM (red), with conventional paraffin PCM (blue), with the hybrid graphene-PCM at 1 wt. % loading (orange) and with the hybrid graphene-PCM at 20 wt. % loading (green). Note that the developed hybrid graphene-PCM strongly reduces the temperature rise inside the battery by simultaneously absorbing the heat and conducting it to the outside shell. The reduction in the temperature rise can be made stronger with a proper design of the outside

heat sink.

**Figure 4: Numerical simulation of temperature rise inside the battery pack.** Simulated temperature profiles in Li-ion battery packs obtained using the measured values of the specific heat and thermal conductivity. The simulation data are in agreement with the experiments.

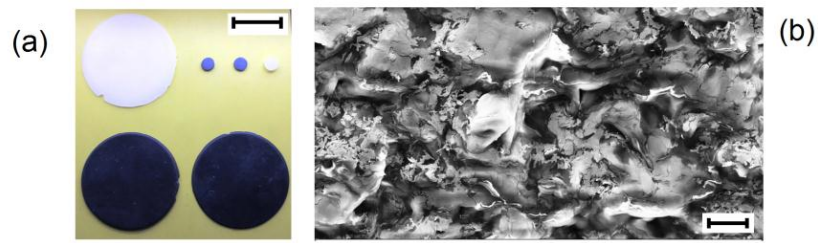

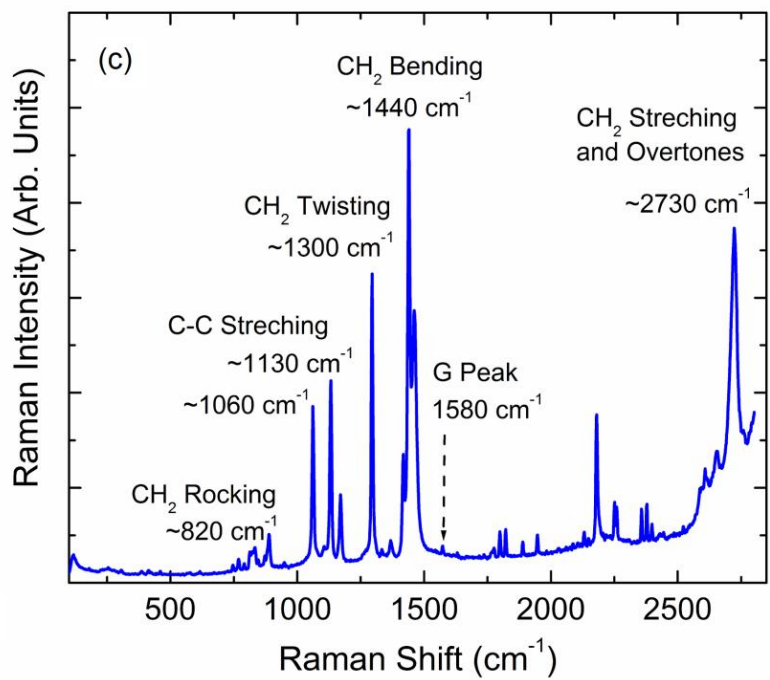

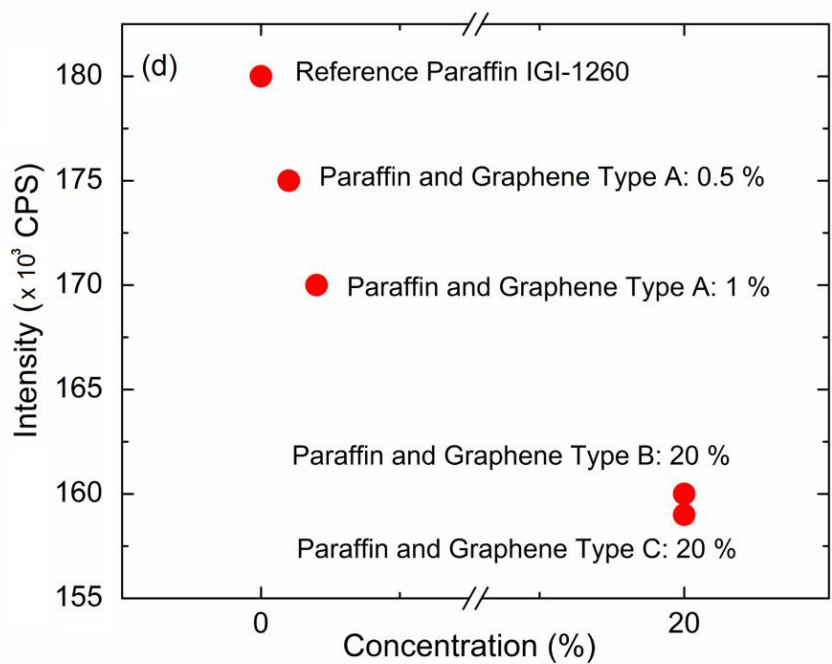

Figure 1

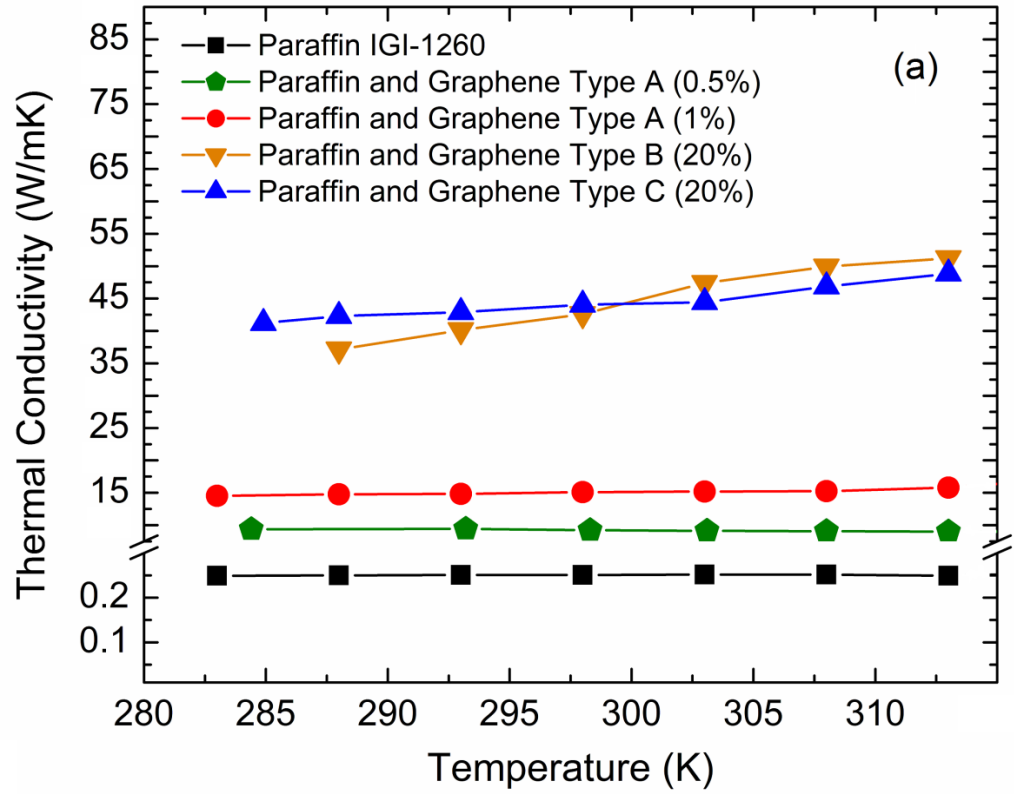

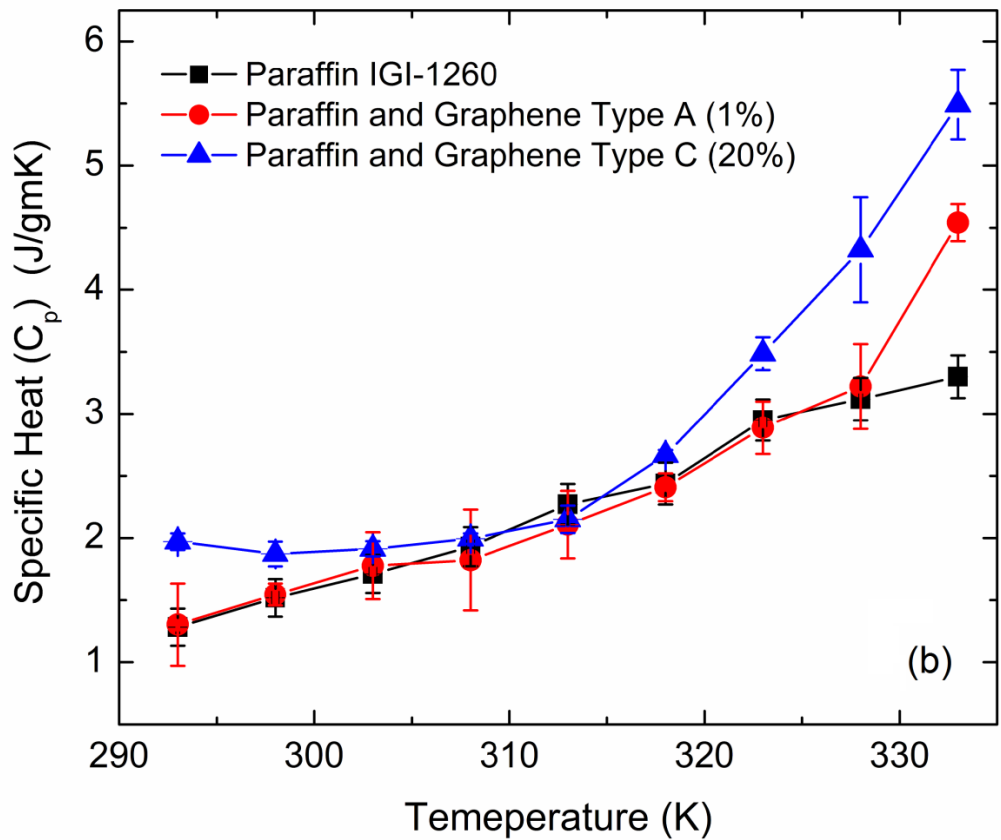

Figure 2

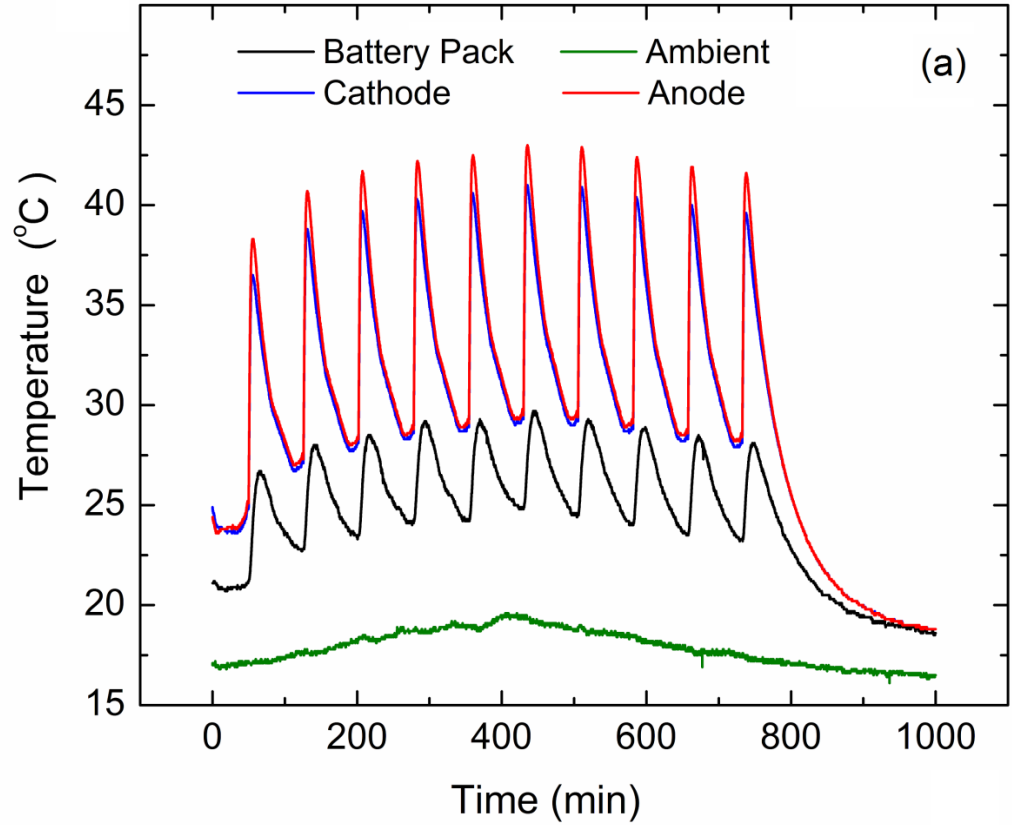

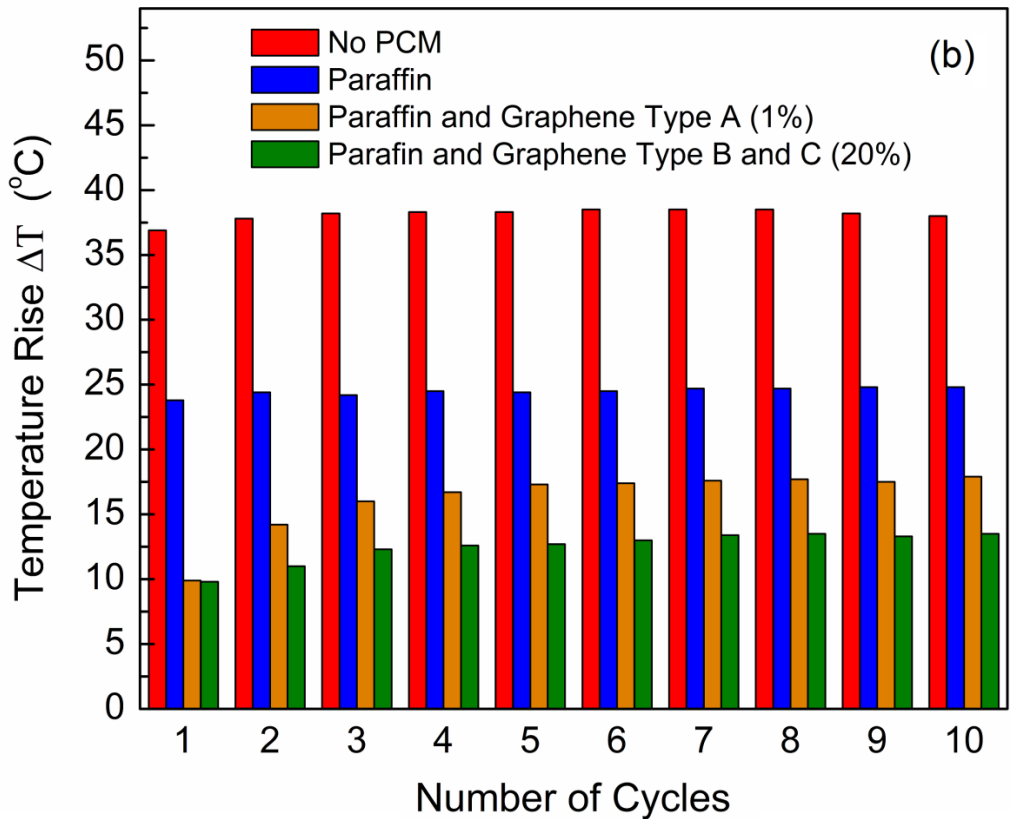

Figure 3

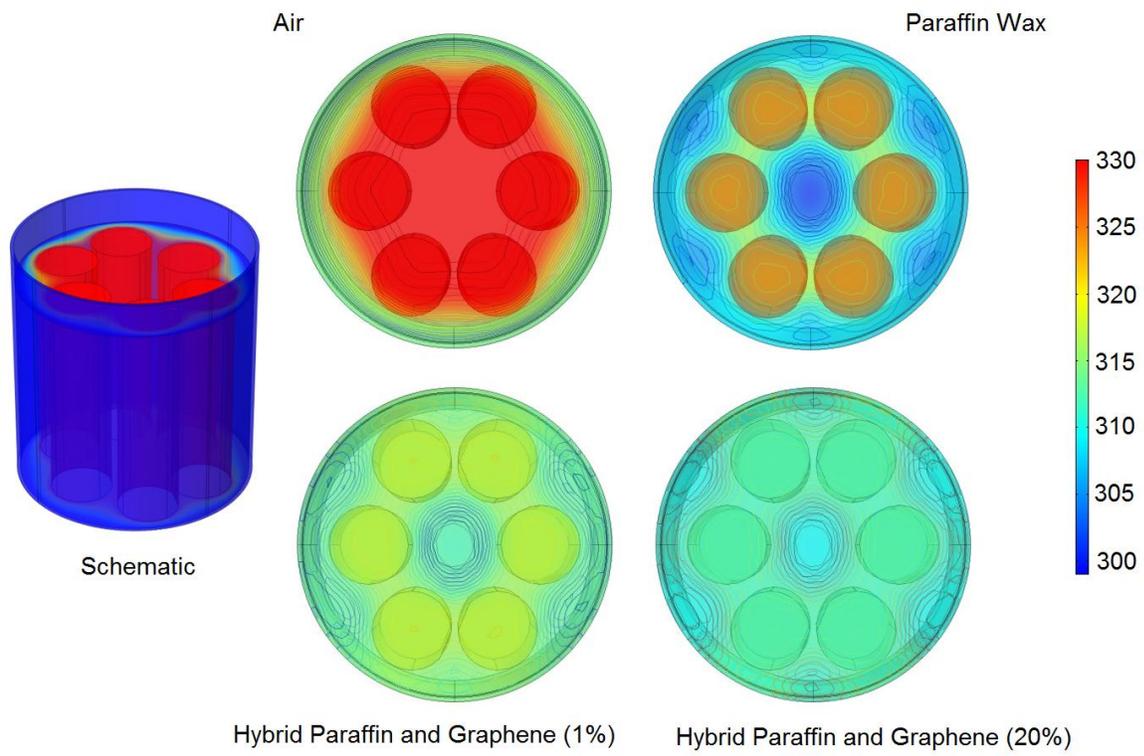

Air

Paraffin Wax

Schematic

Hybrid Paraffin and Graphene (1%)    Hybrid Paraffin and Graphene (20%)

330
325
320
315
310
305
300

Figure 4